\documentclass[preprintnumbers,floatfix,12pt,superscriptaddress,nofootinbib]{revtex4}

\usepackage{amsfonts}
\usepackage{amssymb,epsfig}
\usepackage{latexsym}
\usepackage{amsmath}
\usepackage{graphicx}

\begin{document}

\title{Statefinder diagnostic and $w-w^{\prime}$ analysis for interacting polytropic gas dark energy model.}
\author{M. Malekjani$^{1}$ \footnote{%
Email: \text{malekjani@basu.ac.ir}}, A. Khodam-Mohammadi$^{1}$ \footnote{%
Email: \text{khodam@basu.ac.ir}}} \affiliation{Department of
Physics, Faculty of Science, Bu-Ali Sina University, Hamedan 65178,
Iran\\}

\begin{abstract}
\vspace*{1.5cm} \centerline{\bf Abstract} \vspace*{1cm} The
interacting polytropic gas dark energy model is investigated from
the viewpoint of statefinder diagnostic tool and $w-w^{\prime}$
analysis. The dependency of the statefinder parameters on the
parameter of the model as well as the interaction parameter between
dark matter and dark energy is calculated. We show that different
values of the parameters of model and different values of
interaction parameter result different evolutionary trajectories in
$s-r$ and $w-w^{\prime}$ planes. The polytropic gas model of dark
energy mimics the standard $\Lambda$CDM model at the early time.
\end{abstract}
\maketitle

\newpage
\section{Introduction}
Recent astronomical data obtained by SNe Ia \cite{c1}, WMAP
\cite{c2}, SDSS \cite{c3} and X-ray \cite{c4} experiments suggest
that our universe expands under an accelerated expansion. In the
framework of standard cosmology, a dark energy component with
negative pressure is responsible for this acceleration. A major
puzzle of cosmology is the nature of dark energy and therefore many
theoretical models have been proposed to interpret the behavior of
dark energy. Although the earliest and simplest model is the
Einstein's cosmological constant, but it suffers from tow deep
theoretical problems namely the "fine-tuning" and "cosmic
coincidence". The other models for dark energy scenario are the
dynamical models in which the EoS parameter is time-varying.
According to some analysis on the SNe Ia observational data, it has
been shown that the time-varying dark energy models give a better
fit compare with a cosmological constant \cite{c44}. The dynamical
dark energy models are classified in tow different categories:
 (\emph{i}) The scalar fields
including quintessence \cite{c5}, phantom \cite{c6}, quintom
\cite{c7}, K-essence \cite{c8}, tachyon \cite{c9}, dilaton
\cite{c10} and so forth. (\emph{ii}) The interacting models of dark
energy such as Chaplygin gas models \cite{c100, setare22},
braneworld models \cite{101}, holographic \cite{c11} and agegraphic
\cite{c12} models. The interacting dark energy models have also been
investigated in \cite{jamil2}.

 The holographic dark energy model comes from the holographic principle of quantum gravity \cite{c13} and the
agegraphic model has been proposed based on the uncertainty relation
of quantum mechanics together with general relativity \cite{c14}.
Recent observational data gathered from the Abell Cluster A586
support the interaction between dark matter and dark energy
\cite{c15}. However the strength of this interaction is not clearly
identified \cite{c16}. \\
Here in this work we focus on the polytropic gas model as a dark
energy model to explain the cosmic acceleration. In stellar
astrophysics,  the polytropic gas model can explain the equation of
state of degenerate white dwarfs, neutron stars and also the
equation of state of main sequence stars \cite{c19}. The idea of
dark energy with polytropic gas equation of state has been
investigated by U. Mukhopadhyay and S. Ray in cosmology
\cite{ray05}. Recently, Karami et al. investigated the interaction
between dark energy and dark matter in polytropic gas scenario, the
phantom behavior of polytropic gas, reconstruction of $f(T)$-
gravity from the polytropic gas and the correspondence between
polytropic gas and agegraphic dark energy model
\cite{c17,c18,karam20}. The cosmological implications of polytropic
gas dark energy model is also discussed in \cite{malek1}. The
evolution of deceleration parameter in the context of polytropic gas
dark energy model represents the decelerated expansion at the early
universe and accelerated phase later. Depending on the parameters of
the model, the polytropic gas can achieve a quintessence regime. The
potential and the dynamics of K-essence, dilaton and tachyon fields
according to the evolution of polytropic gas model is also
investigated in \cite{malek1}. \\The  polytropic gas is a
phenomenological model of dark energy. In a phenomenological model,
the pressure $p$ is a function of energy density $\rho$, i.e.,
$p=-\rho-f(\rho)$ \cite{c177}. For $f(\rho)=0$, the equation of
state of phenomenological models can cross $w=-1$, i.e., the
cosmological constant model. Nojiri, et al. investigated four types
 singularities for some illustrative examples of phenomenological
 models \cite{c177}. The polytropic gas model has a type III.
 singularity in which the singularity takes place at a
 characteristic scale factor $a_s$.\\
Since many theoretical dark energy models have been proposed to
explain the accelerated expansion of the universe, therefore a
sensitive test which can discriminate between these models is
required. The Hubble parameter, $H=\dot{a}/a$, (first time
derivative) and the deceleration parameter $q=-\ddot{a}/aH^2$
(second time derivative) are the geometrical parameters to describe
the expansion history of the universe. Since $\dot{a}>0$, hence
$H>0$ means the expansion of the universe. Also $\ddot{a}>0$, i.e.
$q<0$, indicates the accelerated expansion of the universe. Since
the various dark energy models give $H>0$, $q<0$ at the percent
time, hence the Hubble parameter and deceleration parameter can not
discriminate dark energy models. For this aim we need a higher order
of time derivative of scale factor. Sahni et al. \cite{sah03} and
Alam et al. \cite{alam03}, by using the third time derivative of
scale factor, introduced the statefinder pair \{s,r\} in order to
remove the degeneracy of $H$ and $q$ at the present time. The
statefinder pair has been given by
\begin{equation}\label{state1}
r=\frac{\dddot{a}}{aH^3},s=\frac{r-1}{3(q-1/2)}
\end{equation}
Depending the statefinder diagnostic tool on the scale factor
indicates that the statefinder parameters are geometrical. Up to
now, the various dark energy models have been studied from the
viewpoint of statefinder diagnostic. The various dark energy models
have different evolutionary trajectories in \{s, r\} plane,
therefore the statefinder tool can discriminate these models. The
well known $\Lambda$CDM model is related to the fixed point
\{s=0,r=1\} in the $s-r$ plane \cite{sah03}. The other dynamical
dark energy models that have been investigated by statefinder
diagnostic tool are:\\
the quintessence DE model \cite{sah03, alam03} , the interacting
quintessence models \cite{zim, zhang055}, the holographic dark
energy models \cite{zhang056, zhang057} , the holographic dark
energy model in non-flat universe \cite{setar}, the phantom model
\cite{chang}, the tachyon \cite{shao}, the generalized chaplygin gas
model \cite{malek11}, the interacting new agegraphic DE model in
flat and non-flat universe  \cite{ zhang058, malek12}, the
agegraphic dark energy model with and without interaction in flat
and non-flat universe \cite{wei077, malek13} are analyzed through
the statefinder diagnostic tool. Recently, the statefinder parameters have been investigated by considering
the variable gravitational constant $G$ \cite{jamil3}.  \\
In addition to statefinder diagnostic, the other analysis to
discriminate between dark energy models is $w-w^{\prime}$ analysis
that have been used widely in the papers \cite{malek12, wei077,
malek13,che06, chib06, berg06, zhao06, zhao07, lide06, gu06, hua07a,
hua07b, linde07}. In this work we investigate the interacting
polytropic gas model by statefinder diagnostic tool and
$w-w^{\prime}$ analysis. We introduce the interacting polytropic gas
dark energy model in sect. II. The numerical results is presented in
sect. III. We conclude in sect.IV.

\section{Interacting polytropic gas dark energy model}
In this section we give a brief review of the polytropic gas model
for dark energy scenario. For more details and discussion see
\cite{malek1}. The equation of state (EoS) of polytropic gas is
given by
\begin{equation}\label{poly}
p_{\Lambda}=K\rho_{\Lambda}^{1+\frac{1}{n}},
\end{equation}
where $K$ and $n$ are the polytropic constant and polytropic index,
respectively \cite{c19}.\\In the non-flat FRW universe including
dark energy and dark matter components, the Friedmann equation is
given by
\begin{equation}\label{frid1}
H^{2}+\frac{k}{a^{2}}=\frac{1}{3M_{p}^{2}}(\rho _{m}+\rho
_{\Lambda})
\end{equation}%
where $H$ is the Hubble parameter, $M_p$ is the reduced Planck mass
and $k=1,0,-1$ is a curvature parameter corresponding to the closed,
flat and open universe, respectively. $\rho_m$ and $\rho_{\Lambda}$
are the energy densities of dark matter and dark energy, respectively.\\
The dimensionless energy densities are defined as
\begin{equation}\label{denergy}
\Omega_{m}=\frac{\rho_m}{\rho_c}=\frac{\rho_m}{3M_p^2H^2}, ~~~\\
\Omega_{\Lambda}=\frac{\rho_{\Lambda}}{\rho_c}=\frac{\rho_{\Lambda}}{3M_p^2H^2}~~\\
\Omega_k=\frac{k}{a^2H^2}
\end{equation}
Therefore the Friedmann equation (\ref{frid1}) can be written as
\begin{equation}
\Omega _{m}+\Omega _{\Lambda}=1+\Omega _{k}.  \label{Freq2}
\end{equation}%
 In a universe dominated by interacting dark energy and
dark matter, the total energy density, $\rho=\rho_m+\rho_{\Lambda}$,
satisfies a conservation equation
\begin{equation}
\dot{\rho}+3H(\rho+p)=0
\end{equation}
However, by considering the interaction between dark energy and dark
matter, the energy density of dark energy and dark matter does not
conserve separately and in this case the conservation equations for
each component are given by
\begin{eqnarray}
\dot{\rho _{m}}+3H\rho _{m}=Q, \label{contm}\\
\dot{\rho _{\Lambda}}+3H(\rho_{\Lambda}+p_{\Lambda})=-Q,
\label{contd}
\end{eqnarray}%
where  $Q$ represents the interaction between dark components and
can be given as one of the following forms \cite{c21}
%\begin{equation}
%Q=3 \alpha H\rho_{\Lambda}, ~~~ 3\beta H\rho_m, ~~~ 3\gamma
%H(\rho_{\Lambda}+\rho_m),
%\end{equation}
\begin{equation}
Q=\left\{
\begin{array}{ll}
3 \alpha H\rho_{\Lambda} \\
\vspace{0.75mm} 3\beta H\rho_m \\
3\gamma H(\rho_{\Lambda}+\rho_m)%
\end{array}%
\right.   \label{Q-term}
\end{equation}
where  $\alpha$, $\beta$ and $\gamma$ are the dimensionless
constants. The Hubble parameter $H$ in the $Q$-terms is considered
for mathematical simplicity. The interaction parameter $Q$ indicates
the decay rate of dark energy to the dark matter component. The
interaction between dark energy and dark matter is also studied in
\cite{c233}. Here same as our previous works, for mathematical
simplicity, we consider the first form of interaction parameter $Q$
\cite{malek12, malek13, malek1}.  Using Eq.(\ref{poly}), the
integration of
 continuity equation for interacting dark energy component, i.e. Eq.(\ref{contd}),
 gives
 \begin{equation}\label{rho1}
 \rho_{\Lambda}=\left(\frac{1}{Ba^{\frac{3(1+\alpha)}{n}}-\widetilde{K}}\right)^n,
 \end{equation}
where $B$ is the integration constant,
$\widetilde{K}=\frac{K}{1+\alpha}$ and $a$ is the scale factor. A
positive energy density for any value of  $n$ is achieved when
$Ba^{3(1+\alpha)/n}>\widetilde{K}$.  In the case of even $n$, we
have positive energy density for any condition of $\widetilde{K}$.
The phantom behavior of interacting polytropic gas dark energy has
been studied in \cite{c17}. In the case of
$Ba^{3(1+\alpha)/n}=\widetilde{K}$, we have $\rho\rightarrow \infty$
and the polytropic gas has a finite-time singularity at
$a_c=(\widetilde{K}/B)^{n/3(1+\alpha)}$. This type of singularity,
in which at a characteristic scale factor $a_s$, the energy density
$\rho\rightarrow\infty$ and the pressure density
$|p|\rightarrow\infty$, has been indicated by type III singularity
\cite{c177}.\\
Substituting $Q=3\alpha H\rho_{\Lambda}$ in (\ref{contd}) obtains
%\begin{equation}\label{contm2}
%\dot{\rho _{m}}+3H(1+w_m^{eff})\rho _{m}=0
%\end{equation}
\begin{equation}\label{contd2}
\dot{\rho _{\Lambda}}+3H(1+\alpha+w_{\Lambda})\rho_{\Lambda}=0,
\end{equation}
%where we define the effective EoS parameters as:
% $w_m^{eff}=-b^2\frac{1+\Omega_k}{\Omega_m}$ and $w_{\Lambda}^{eff}=w_{\Lambda}+b^2\frac{1+\Omega_k}{\Omega_{\Lambda}}$.\\
Taking the derivative of Eq.(\ref{rho1}) with respect to time, one
can obtain
\begin{equation}\label{dotrho}
\dot{\rho_{\Lambda}}=-3BH(1+\alpha)a^{\frac{3(1+\alpha)}{n}}\rho_{\Lambda}^{1+\frac{1}{n}}
\end{equation}
Substituting Eq.(\ref{dotrho}) in (\ref{contd2}) and using
Eq.(\ref{rho1}) , we can obtain the  EoS parameter of interacting
polytropic gas as
\begin{equation}\label{eos1}
 w_{\Lambda}=-1-\frac{a^{\frac{3(1+\alpha)}{n}}}{c-a^{\frac{3(1+\alpha)}{n}}}-\alpha
\end{equation}
 where $c=\widetilde{K}/B$. We see that the interacting polytropic gas model behaves as a
phantom model, i.e. $w_{\Lambda}<-1$, when $c>a^{3(1+\alpha)/n}$. It
is also clear to see that at the early time ($a\rightarrow 0$) and
the absence of interaction between dark matter and dark energy
($\alpha=0$), the polytropic gas mimics the constant, i.e.
$w_{\Lambda}\rightarrow -1$. The evolution of EoS parameter can be
obtained by differentiating of (\ref{eos1}) as follows
\begin{equation}\label{eosp}
w_{\Lambda}^{\prime}=-\frac{3(1+\alpha)ca^{\frac{3(1+\alpha)}{n}}}{n(c-a^{\frac{3(1+\alpha)}{n}})^2}
\end{equation}
where prime denotes the derivative with respect to $x=\ln{a}$. Using
Eqs.(\ref{rho1}) and (\ref{denergy}) the density parameter of
interacting polytropic gas is given by
\begin{equation}\label{denergy2}
\Omega_{\Lambda}=\frac{(Ba^{\frac{3(1+\alpha)}{n}}-\widetilde{K})^{-n}}{3M_p^2H^2}
\end{equation}
Taking the time derivative of Eq.(\ref{denergy2})  and using
$\Omega^{\prime}=\dot{\Omega}/H$, yields
\begin{equation}\label{motion}
\Omega_{\Lambda}^{\prime}=-\Omega_{\Lambda}\Big(\frac{3(1+\alpha)a^{\frac{3(1+\alpha)}{n}}}{a^{\frac{3(1+\alpha)}{n}}-c}+2\frac{\dot{H}}{H^2}\Big)
\end{equation}
Taking the time derivative of Friedmann equation (\ref{frid1}) and
using Eqs.(\ref{rho1}), (\ref{Freq2}), (\ref{contm}),
(\ref{denergy2}) and $Q=3\alpha H\rho_{\Lambda}$, one can find that
\begin{equation}\label{doth}
\frac{\dot{H}}{H^2}=-\frac{3}{2}\Big[\Omega_{\Lambda}\frac{c(1+\alpha)}{a^\frac{3(1+\alpha)}{n}-c}+1+\frac{\Omega_k}{3}\Big]
\end{equation}
Substituting this relation into Eq.(\ref{motion}), we obtain the
evolutionary equation for energy density parameter of interacting
polytropic gas as:
\begin{equation}\label{omega_evol}
\Omega_{\Lambda}^{\prime}=-3\Omega_{\Lambda}\Big[\frac{c}{a^{\frac{3(1+\alpha)}{n}}-c}(1-\Omega_{\Lambda})+\alpha
\frac{a^{\frac{3(1+\alpha)}{n}}-c\Omega_{\Lambda}}{a^{\frac{3(1+\alpha)}{n}}-c}-\frac{\Omega_k}{3}\Big],
\end{equation}
The evolution of density parameter $\Omega_{\Lambda}$ has been
discussed in \cite{malek1}. It has been shown that the polytropic
gas dark energy model can describe the matter dominated universe at
the early time, $\Omega_{\Lambda}\rightarrow 0$, and dark energy
universe at the late time, $\Omega_{\Lambda}\rightarrow 1$, see
Fig.(2) of \cite{malek1}.\\
 The deceleration parameter is given by
\begin{equation}\label{qdece}
q=-\frac{\ddot{a}}{aH^2}=-1-\frac{\dot{H}}{H^2}
\end{equation}
 Substituting (\ref{doth}) in (\ref{qdece}), the deceleration parameter
 can be obtained as
\begin{equation}\label{qdece2}
q=-1+\frac{3}{2}\Big[\Omega_{\Lambda}\frac{c(1+\alpha)}{a^\frac{3(1+\alpha)}{n}-c}+1+\frac{\Omega_k}{3}\Big]
\end{equation}
The evolution of $q$ for interacting polytropic gas model has also
been presented in \cite{malek1}. At the early time the universe has
a decelerated expansion, $q>0$, and enters into the accelerated
phase later, $q<0$, see Fig.(3) of \cite{malek1}.\\
Using (\ref{eos1}), (\ref{doth}) and (\ref{qdece2}), we have
\begin{eqnarray}\label{ddoth1}
&&\frac{\ddot{H}}{H^3}=-\frac{9}{2}\Omega_{\Lambda}(1+\alpha)(\alpha+w_{\Lambda})[(1+\alpha)(-w_{\Lambda}+\Omega_{\Lambda}\alpha+\Omega_{\Lambda}w_{\Lambda})-\alpha(\alpha+2)]\nonumber \\
&&-\frac{3}{2}\Omega_{\Lambda}(1+\alpha)w_{\Lambda}^{\prime}+\frac{9}{2}[\Omega_{\Lambda}(1+\alpha)(\alpha+w_{\Lambda})+1]^2
\end{eqnarray}

At what follows, we derive the statefinder parameters (${s,r}$) for
polytropic gas model in the interacting spatially flat universe.
Using the definition of statefinder parameters in (\ref{state1}),
one can obtain
\begin{equation}\label{r1}
r=\frac{\dddot{a}}{aH^3}=\frac{\ddot{H}}{H^3}-3q-2
\end{equation}
Inserting (\ref{qdece2}) and (\ref{ddoth1}) in (\ref{r1}) and using
(\ref{motion}) we have
\begin{equation}\label{r2}
r=1+\frac{3}{2}\Omega_{\Lambda}(1+\alpha)[3(1+\alpha)(\alpha+w_{\Lambda})(1+\alpha+w_{\Lambda})-w_{\Lambda}^{\prime}]
\end{equation}
Inserting (\ref{qdece2}) and (\ref{r2}) in (\ref{state1}), the
parameter $s$ for interacting polytropic gas is obtained as
\begin{equation}\label{s3}
s=\frac{2}{3}\frac{3\alpha(\alpha+1)^2+3\alpha
w_{\Lambda}(2\alpha+w_{\Lambda}+3)+3w_{\Lambda}(1+w_{\Lambda})-w_{\Lambda}^{\prime}}{\alpha+w_{\Lambda}}
\end{equation}
In the limiting case of $w_{\Lambda}=-1$, it is obvious
$w_{\Lambda}^{\prime}=0$ and in the absence of interaction between
dark matter and dark energy, i.e. $\alpha=0$, the statefinder
parameters reduce to $\{s=0, r=1\}$ which is coincide to the
location of standard $\Lambda CDM$ model in $s-r$ plane.

\section{Numerical results}
In the present section we give the numerically description of the
evolutionary trajectories of the statefinder parameters in $s-r$
plane for interacting polytropic gas dark energy model in the flat
universe. We also perform the $w-w^{\prime}$ analysis for this
model. Here we set $\Omega_m^0=0.3$ and $\Omega_{\Lambda}^0=0.7$ for
the density parameters of dark matter and dark energy at the present time, respectively. \\
\subsection{Statefinder diagnostic}
The statefinder pair $\{s,r\}$ in this model is given by (\ref{r2})
and (\ref{s3}). One can easily see the dependency of the $\{s,r\}$
on the EoS parameter , $w$, as well as the interaction parameter
$\alpha$ in (\ref{r2}) and (\ref{s3}). In the limiting case of
non-interacting polytropic gas ($\alpha=0$), from (\ref{eos1}) we
see that at the early time ($a\rightarrow 0$) the EoS parameter
$w_{\Lambda}\rightarrow -1$. From (\ref{eosp}), we also have
$w_{\Lambda}^{\prime}\rightarrow 0$. From (\ref{r2}) and (\ref{s3})
we see that at the early time the statefinder parameters for
non-interacting flat universe are ($s=0, r=1$) which is coincide to
the location of spatially flat $\Lambda CDM$ model in the $s-r$
plane. Hence, the polytropic gas model mimics the $\Lambda CDM$
model at the early time.\\
In Fig.(1), the evolutionary trajectories of interacting polytropic
gas model is plotted for different values of interaction parameter
$\alpha$. Here we fix the parameters of the model as $c=2, n=4$. The
standard $\Lambda CDM$ fixed point is indicated by star symbol in
this diagram. The colored circles on the curves show the present
values of statefindr pair $\{s_0, r_0\}$. Different values of
$\alpha$ result different evolutionary trajectories in $s-r$ plane.
Hence the interaction parameter can influence on the evolutionary
trajectory of polytropic gas model in $s-r$ plane. For larger value
of $\alpha$, the present value $s_0$ decreases and the present value
$r_0$ increases. The distance of the point ($s_0, r_0$) form the
$\Lambda CDM$ fixed point (i.e. $s=0,r=1$) becomes larger by
increasing the interaction parameter $\alpha$. While the universe
expands, the evolutionary trajectory of interacting polytropic gas
dark energy model evolves from the $\Lambda CDM$ at the early time,
then $r$ increases and $s$ decreases. The present value $\{s_0,
r_0\}$ are valuable, if it can be extracted from the future data of
SNAP (SuperNova Acceleration Probe) experiments. Therefore, the
statefinder diagnostic tool with future SNAP observation are useful
to discriminate between various dark energy models.\\
In Fig.(2), the evolutionary trajectories for interacting polytropic
gas are plotted for different values of the parameters of the model.
Here we fix the interaction parameter as $\alpha=0$. In left panel,
the parameter $n$ is fixed and the parameter $c$ is varied.
Different values of $c$ gives the different evolutionary
trajectories in $s-r$ plane. Therefore the parameter $c$ of the
model can affect on the evolutionary trajectories in $s-r$ plane.
Like Fig.(1), the present value of statefinder pair, i.e.
$\{s_0,r_0\}$ is indicated by colore circles on the curves. For
larger values of $c$, $r_0$ decreases and $s_0$ increases. The
distance of the point ($s_0,r_0$) to the location of standard
$\Lambda$CDM fixed point becomes shorter for larger value of $c$. In
right panel the parameter $c$ is fixed and the parameter $n$ is
varied. Same as left panel, the interaction parameter is fixed to
$\alpha=0$. Here we also see that different values of $n$ gives
different evolutionary trajectories in $s-r$ plane. For larger
values of $n$, we see $r_0$ decreases and $s_0$ increases. Here we
see that, same as parameter $c$, the distance of the point
($s_0,r_0$) to the location of standard $\Lambda$CDM fixed point
becomes shorter for larger value of $n$.\\
\subsection{$w_{\Lambda}-w_{\Lambda}^{\prime}$ analysis}
In addition to the statefinder diagnostic, another analysis to
discriminate various models of dark energy is
$w-w_{\Lambda}^{\prime}$ analysis. Here we apply this analysis for
interacting polytropic gas dark energy model. In this analysis the
standard $\Lambda CDM$ model corresponds to $\{w_{\Lambda}=-1,
w_{\Lambda}^{\prime}=0\}$. The evolution of $w_{\Lambda}$ and
$w_{\Lambda}^{\prime}$ are given by (\ref{eos1}) and (\ref{eosp}),
respectively. In Fig.(3), the evolutionary trajectories of
interacting polytropic gas dark energy for different values of
interaction parameter are shown in
$w_{\Lambda}-w_{\Lambda}^{\prime}$ plane. Here we fix the parameter
of the polytropic model as $c=2$ and $n=4$. One can see the
different values of $\alpha$ result different trajectories in
$w_{\Lambda}-w_{\Lambda}^{\prime}$ plane. The present value
$w_{\Lambda}^{0}-w_{\Lambda}^{\prime 0}$ is dependent on the
interaction parameter. Larger value of $\alpha$ obtains smaller
values of $w_{\Lambda}^{0}$ and $w_{\Lambda}^{\prime 0}$.\\
In Fig.(4), the evolutionary trajectories are plotted in the absence
of interaction parameter, i.e. $\alpha=0$. Here we perform the
$w_{\Lambda}-w_{\Lambda}^{\prime}$ analysis for different values of
the parameters of model. In left panel the parameter $n$ is fixed
and in the right panel the parameter $c$ is fixed. In these
diagrams, by expanding the universe, the evolutionary trajectories
start from the fixed point $w_{\Lambda}=-1, w_{\Lambda}^{\prime}=0$
(i.e. the location of $\Lambda CDM$ fixed point). In left panel, one
can see that the parameter $c$ can only affect the present value
$w_{\Lambda}^{0}-w_{\Lambda}^{\prime 0}$. Different values of $c$
result the same evolutionary trajectory in
$w_{\Lambda}-w_{\Lambda}^{\prime}$ plane. The distance of the
present value $w_{\Lambda}^{0}-w_{\Lambda}^{\prime 0}$ to the
location of $\Lambda CDM$ fixed point (i.e. $\{w_{\Lambda}=-1,
w_{\Lambda}^{\prime}=0\}$) is shorter for larger value of $c$. In
right panel we have different evolutionary trajectories in
$w_{\Lambda}-w_{\Lambda}^{\prime}$ plane for different values of
$n$. The parameter $w_{\Lambda}^{\prime 0}$ increases for larger
value of $n$. Like left panel, the distance of the present value
$w_{\Lambda}^{0}-w_{\Lambda}^{\prime 0}$ to the location of $\Lambda
CDM$ model is shorter for larger value of $n$.

\section{Conclusion}
Since many dynamical dark energy models have been proposed to
interpret the cosmic acceleration, the statefinder diagnostic tool
based on third time derivative of scale factor is given to
discriminate between them. The statefinder diagnostic combined with
future SNAP observation can be useful to discriminate between
various dark energy models. Here we studied the statefinder
diagnostic tool for interacting polytropic gas model in spatially
flat universe. We derive the statefinder parameters $s$ and $r$ in
this model and investigate the dependency of the evolutionary
trajectories in $s-r$ plane on the parameters of the model as well
as the interaction parameter between dark matter and dark energy. We
obtained the present value $\{s_0, r_0\}$ of this model and studied
the dependency of $\{s_0, r_0\}$ on the parameters of the model and
interaction parameter. By expanding the universe, the evolutionary
trajectories start from the $\Lambda CDM$ fixed point then $r$
increases and $s$ decreases. For smaller value of interaction
parameter $\alpha$, the distance of $\{s_0, r_0\}$ from the location
of $\Lambda CDM$ fixed point becomes shorter. Also the larger values
of $c$ and $n$ result the shorter distance from the $\Lambda CDM$
fixed point. The behavior of interacting polytropic gas in $s-r$
plane is similar with the new holographic dark energy model (see
Fig.(4) of \cite{malek15}). For both models, by expanding the
universe, $r$ increases and $s$ decreases. Finally we studied the
$w-w^{\prime}$ analysis for this model. We showed that the
evolutionary trajectories in $w-w^{\prime}$ plane is dependent on
the parameters of the model and also the interaction parameter
$\alpha$. While the universe expands, the trajectories starts from
the location of $\Lambda CDM$ fixed point (i.e. $\{w_{\Lambda}=-1,
w_{\Lambda}^{\prime}=0$\}). Hence the polytropic gas model mimics
the standard $\Lambda$CDM fixed point at the early time. The
agegraphic dark energy model also mimics the $\Lambda$CDM model at
the early time \cite{wei077, malek12}. At future the high-precision
 SNAP-type experiment can be useful to determine the statefinder
parameters precisely and consequently single out the right dark
energy models.

\newpage
\begin{center}
\begin{figure}[!htb]
\includegraphics[width=10cm]{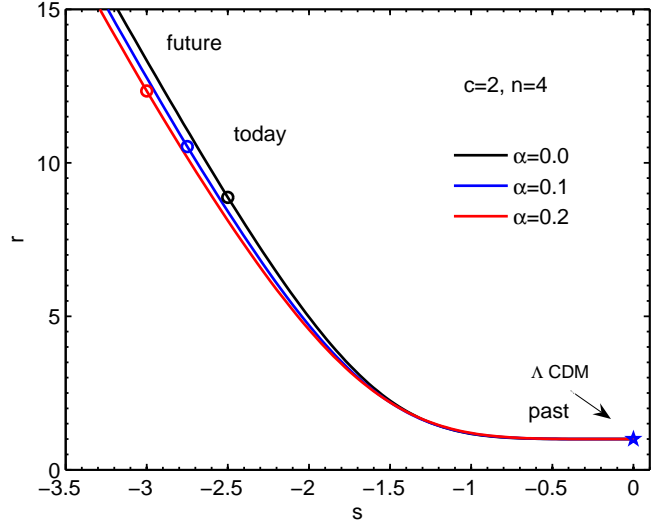}
\caption{The evolutionary trajectories for interacting polytropic
gas model in $s-r$ plane for different values of interaction
parameter $\alpha$. The black curve indicates the non-interacting
case and the blue and red curves represent $\alpha=0.1, 0.2$,
respectively. The circles on the curves show the present value of
the statefinder pair $\{s_0, r_0\}$. The star symbol is related to
the location of standard $\Lambda CDM$ model in $s-r$ plane. The
parameters of the model are chosen as $c=2, n=4$.}
\end{figure}
\end{center}

\newpage

\newpage
\begin{center}
\begin{figure}[!htb]
\includegraphics[width=8cm]{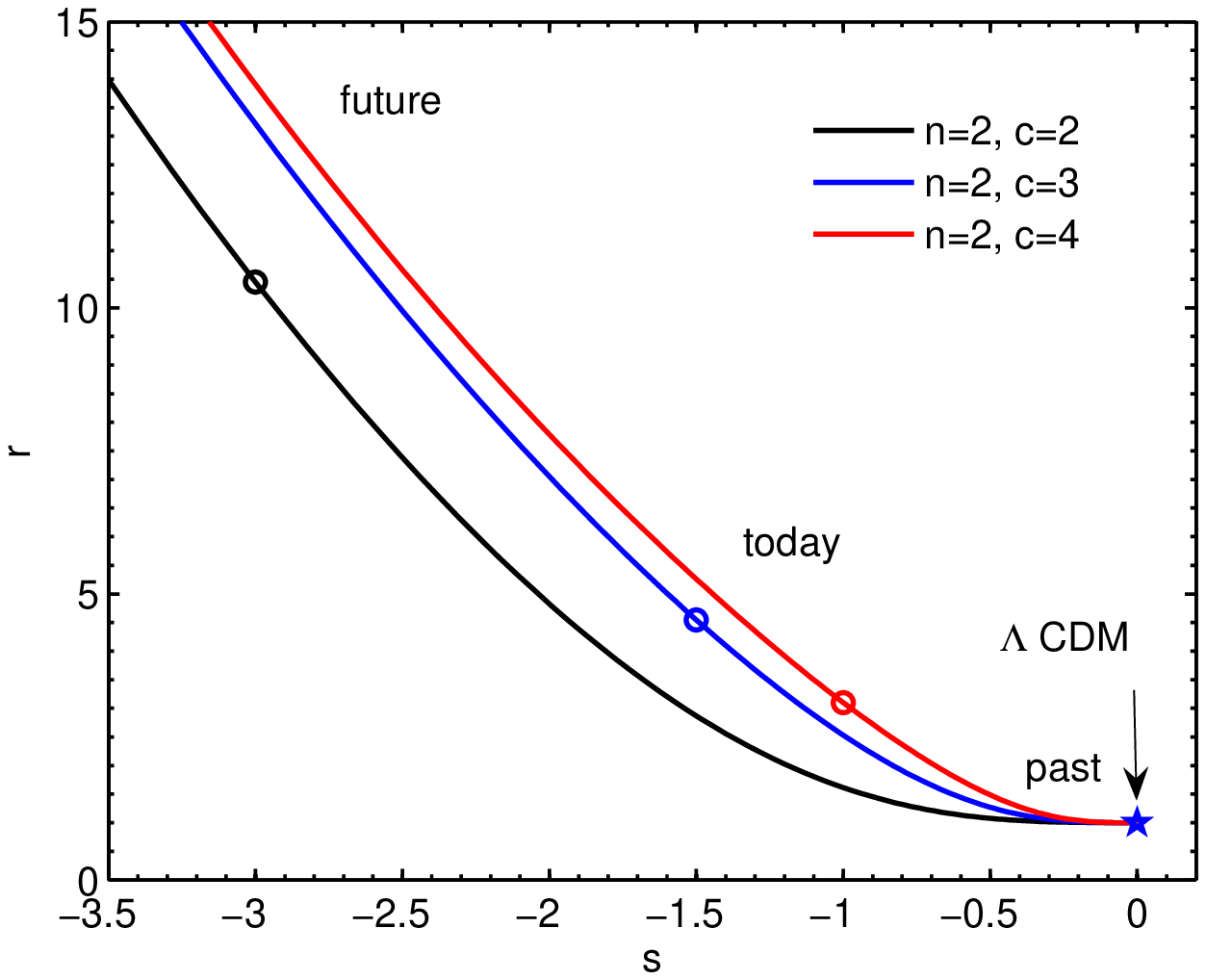}\includegraphics[width=8cm]{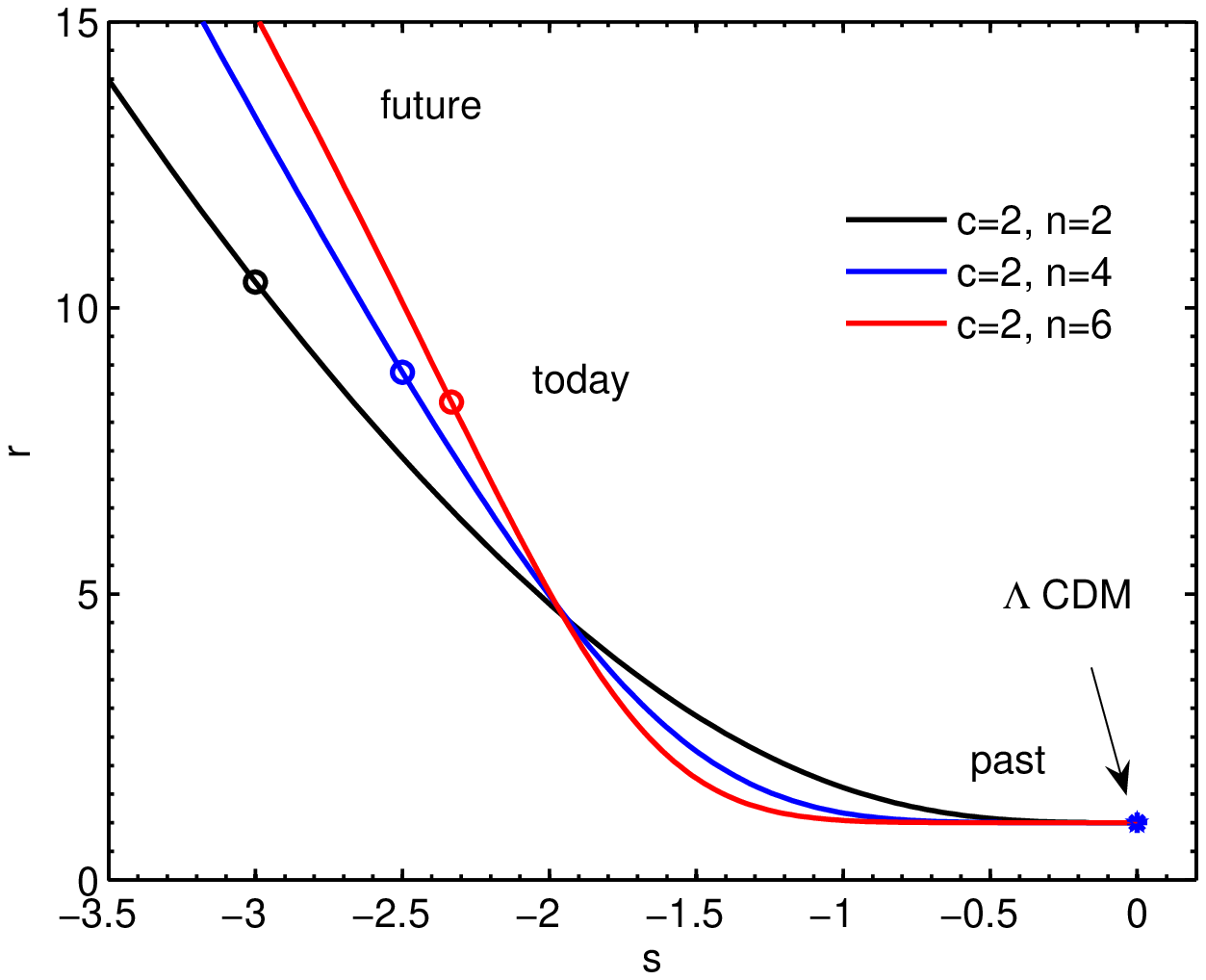}
\caption{The evolutionary trajectories for polytropic gas model in
$s-r$ plane for different illustrative values of parameters $c$ and
$n$. Here we choose the interaction parameter as $\alpha=0$. In left
panel the parameter $n$ is fixed and the parameter $c$ is varied as
$c=2$( black curve ), $c=3$ ( blue curve ), $c=4$ ( red curve ). In
right panel the parameter $c$ is fixed and the parameter $n$ is
varied as $n=2$ ( black curve ), $n=4$ ( blue curve ) and $n=6$ (
red curve ). The circles on the curves show the present value of the
statefinder pair $\{s_0, r_0\}$. The star symbol is related to the
location of standard $\Lambda CDM$ model in $s-r$ plane.}
\end{figure}
\end{center}

\newpage
\begin{center}
\begin{figure}[!htb]
\includegraphics[width=8cm]{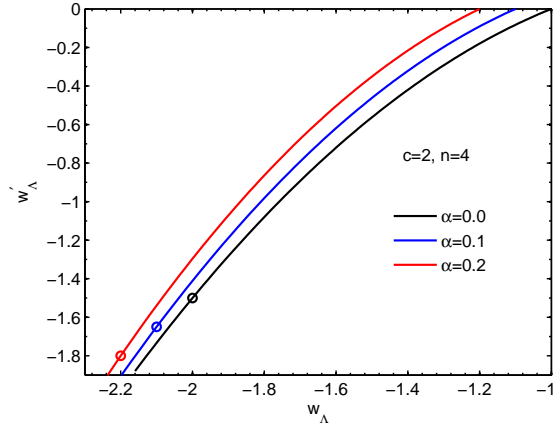}
\caption{The evolutionary trajectories for interacting polytropic
gas model in $w_{\Lambda}-w_{\Lambda}^{\prime}$ plane for different
values of interaction parameter $\alpha$. The black curve shows the
non-interacting case and the blue and red curves indicate
$\alpha=0.1, 0.2$, respectively. The colored circles on the curves
show the present value $w_{\Lambda}^{0}-w_{\Lambda}^{\prime 0}$.
 The parameters of the model are chosen as $c=2, n=4$.}
\end{figure}
\end{center}

\newpage
\begin{center}
\begin{figure}[!htb]
\includegraphics[width=8cm]{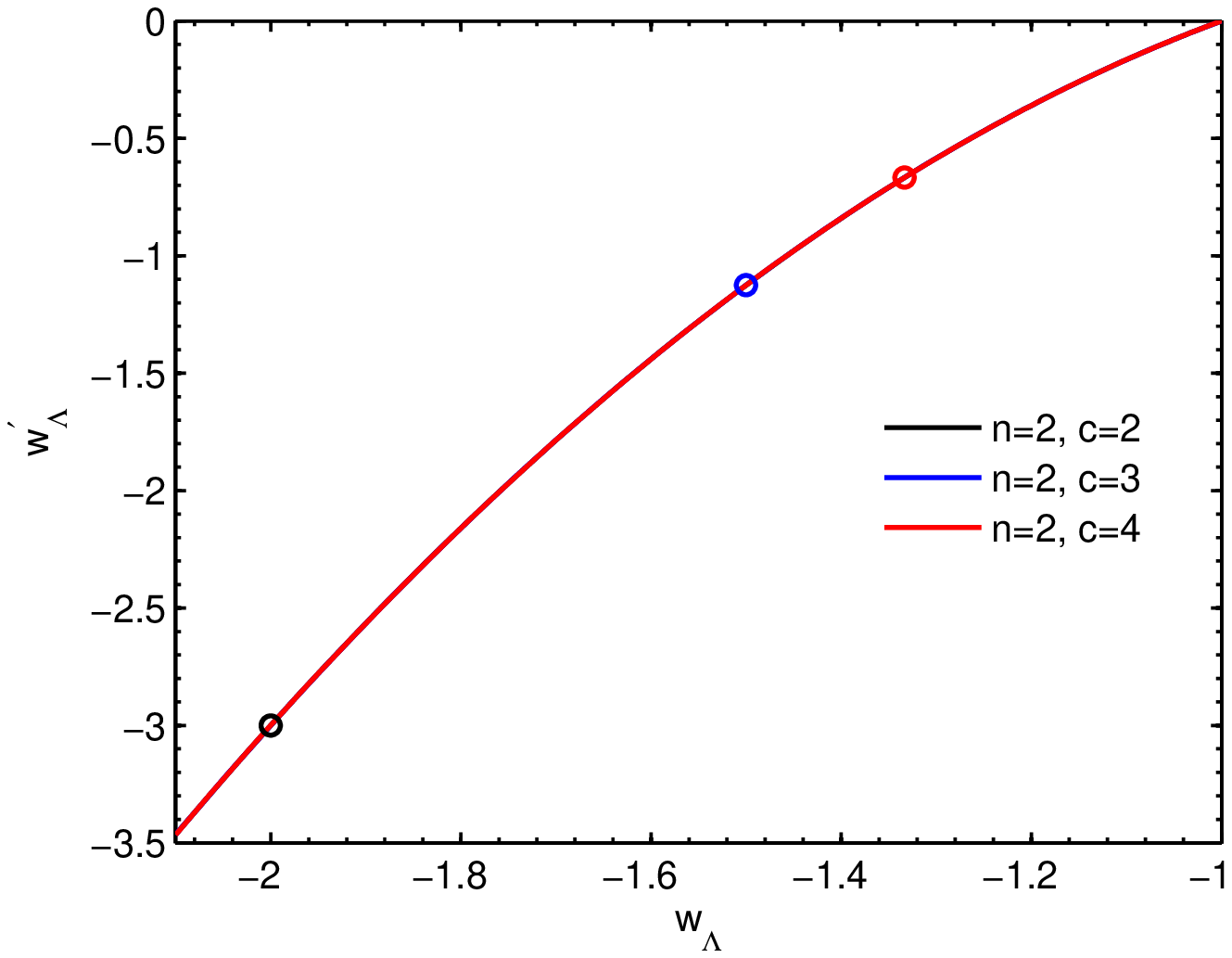}\includegraphics[width=8cm]{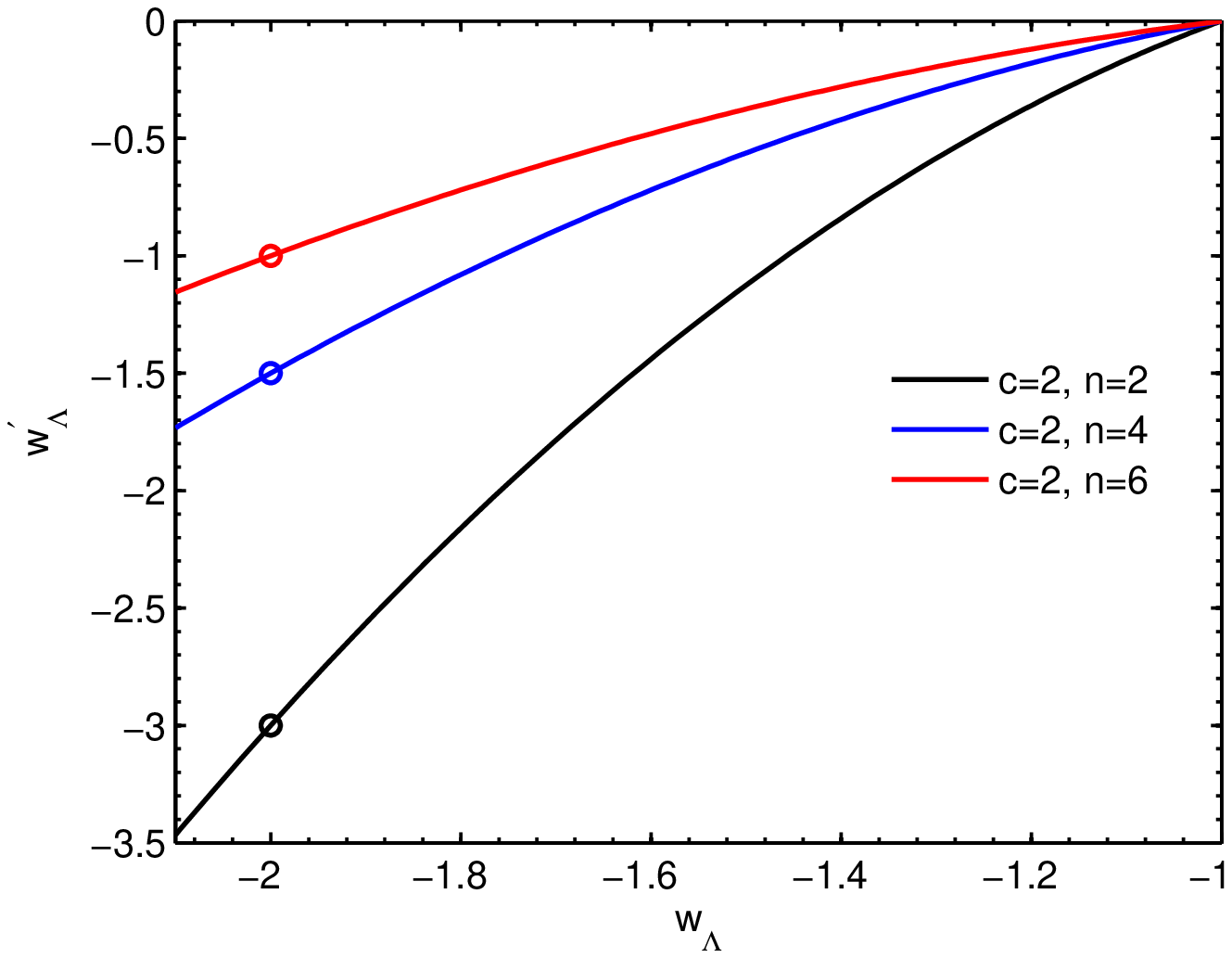}
\caption{The evolutionary trajectories for polytropic gas model in
$w_{\Lambda}-w_{\Lambda}^{\prime}$ plane for different illustrative
values of parameters $c$ and $n$. Here the interaction parameter is
chosen as $\alpha=0$. In left panel the parameter $n$ is fixed and
the parameter $c$ is varied as $c=2$( black curve ), $c=3$ ( blue
curve ), $c=4$ ( red curve ). In right panel the parameter $c$ is
fixed and the parameter $n$ is varied as $n=2$ ( black curve ),
$n=4$ ( blue curve ) and $n=6$ ( red curve ).}
\end{figure}
\end{center}

\end{document}